\documentclass[aps,twocolumn,prl]{revtex4}

\usepackage{graphicx}
\usepackage{revsymb}
\begin{document}

\title{Evidence for reversible control of magnetization in a ferromagnetic material via spin-orbit magnetic field}

\author{Alexander Chernyshov}
\thanks{These authors contributed equally to the project}
\author{Mason Overby}
\thanks{These authors contributed equally to the project}
\affiliation{Department of Physics and Birck Nanotechnology Center, Purdue
University, West Lafayette, Indiana 47907 USA}

\author{Xinyu Liu}
\author{Jacek~K. Furdyna}
\affiliation{Department of Physics, University of Notre Dame, Notre Dame,
Indiana 46556 USA}

\author{Yuli Lyanda-Geller}
\author{Leonid~P. Rokhinson}
%\email[]{leonid@physics.purdue.edu}
%\homepage[]{}
\thanks{To whom correspondence should be addressed. E-mail: leonid@purdue.edu}
\affiliation{Department of Physics and Birck Nanotechnology Center, Purdue
University, West Lafayette, Indiana 47907 USA}

%\date{Submitted on November 19, 2008}
%\date{\today, ---= draft: \jobname.tex =---}
%\pacs{+75.50.Pp, 73.20.Fz, 71.23.-k}

\maketitle \hyphenation{GaMnAs}

{\bf Conventional computer electronics creates a dichotomy between how
information is processed and how it is stored. Silicon chips process
information by controlling the flow of charge through a network of logic gates.
This information is then stored, most commonly, by encoding it in the
orientation of magnetic domains of a computer hard disk. The key obstacle to a
more intimate integration of magnetic materials into devices and circuit
processing information is a lack of efficient means to control their
magnetization. This is usually achieved with an external magnetic field or by
the injection of spin-polarized currents \cite{slonczewski96,berger96,myers99}.
The latter can be significantly enhanced in materials whose ferromagnetic
properties are mediated by charge carriers \cite{chiba04}. Among these
materials, conductors lacking spatial inversion symmetry couple charge currents
to spin by intrinsic spin-orbit (SO)  interactions, inducing nonequilibrium
spin polarization
\cite{aronov89,edelstein90,kalevich90,kato04a,silov04,ganichev08,meier07}
tunable by local electric fields. Here we show that magnetization of a
ferromagnet can be reversibly manipulated by the SO-induced polarization of
carrier spins generated by unpolarized currents. Specifically, we demonstrate
domain rotation and hysteretic switching of magnetization between two
orthogonal easy axes in a model ferromagnetic semiconductor.}

\begin{figure}
\label{sample}
\includegraphics[scale=0.5]{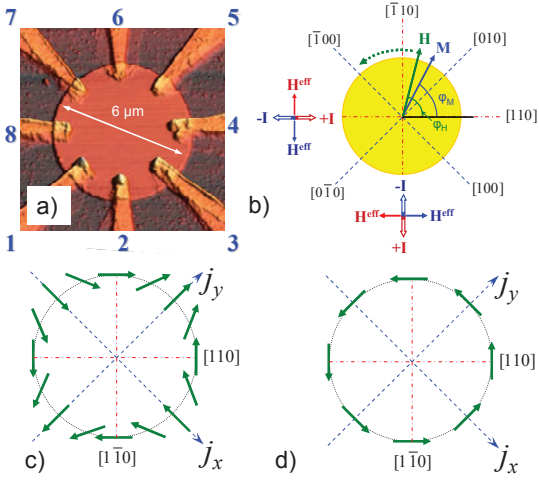}
%\vspace{-1in}
\caption{{\bf Layout of the device and symmetry of the SO fields.}
a) AFM image of sample A with 8 non-magnetic metal contacts. b) Diagram of
device orientation with respect to crystallographic axes, with easy/hard
magnetization axes marked with blue/red dashed lines. Measured directions of
$\mathbf{H^{\mathrm{eff}}}$ field are shown for different current directions.
c,d) Orientation of effective magnetic field with respect to current direction
for c) strain-induced and d) Rashba SO interactions. Current-induced Oersted
field under the contacts has the same symmetry as the Rashba field.}
\end{figure}

In crystalline materials with inversion asymmetry, intrinsic spin-orbit
interactions (SO) couple the electron spin with its momentum $\hbar\mathbf{k}$.
The coupling is given by the Hamiltonian ${\cal
H}_{so}=\frac{\hbar}{2}\mathbf{\hat\sigma\cdot\Omega (k)}$, where $\hbar$ is
the Planck's constant and $\hat\sigma$ is the electron spin operator (for holes
$\hat\sigma$ should be replaced by the total angular momentum $\mathbf{J}$).
Electron states with different sign of the spin projection on
$\mathbf{\Omega(k)}$ are split in energy, analogous to the Zeeman splitting in
an external magnetic field. In zinc-blende crystals such as GaAs there is a
cubic Dresselhaus term\cite{dresselhaus55} $\mathbf{\Omega_D}\propto k^3$,
while strain introduces a term
$\mathbf{\Omega_\varepsilon}=C\Delta\varepsilon(k_x,-k_y,0)$ that is linear in
$k$, where $\Delta\varepsilon$ is the difference between strain in the
$\hat{z}$ and $\hat{x},\hat{y}$ directions\cite{birpikus74}. In wurzite
crystals or in multilayered materials with structural inversion asymmetry there
also exists the Rashba term\cite{bychkov84} $\mathbf{\Omega_R}$ which has a
different symmetry with respect to the direction of $k$, $\mathbf{\Omega_R}=
\alpha_R (-k_y,k_x,0)$, where $\hat z$ is along the axis of reduced symmetry.
In the presence of an electric field the electrons acquire an average
momentum$\hbar\Delta\mathbf{k}(\mathbf{E})$, which leads to the generation of
an electric current $\mathbf{j}=\hat{\rho}^{-1}\mathbf{E}$ in the conductor,
where $\hat{\rho}$ is the resistivity tensor. This current defines the
preferential axis for spin precession
$\langle\mathbf{\Omega}(\mathbf{j})\rangle$. As a result, a nonequilibrium
current-induced spin polarization
$\langle\mathbf{J^E}\rangle\|\langle\mathbf{\Omega(j)}\rangle$ is generated,
whose magnitude $\langle J^E\rangle$ depends on the strength of various
mechanisms of momentum scattering and spin relaxation\cite{aronov89,aronov91}.
This spin polarization has been measured in non-magnetic semiconductors using
optical\cite{kalevich90,kato04a,silov04,golub06,meier07} and electron spin
resonance\cite{wilamowski07} techniques. It is convenient to parameterize
$\langle\mathbf{J^E}\rangle$ in terms of an effective magnetic field
$\mathbf{H^{so}}$. Different contributions to $\mathbf{H^{so}}$ have different
current dependencies ($\propto j\ \mathrm{or}\  j^3$), as well as different
symmetries with respect to the direction of $\mathbf{j}$, as schematically
shown in Fig.~\ref{sample}(c,d), allowing one to distinguish between spin
polarizations in different fields.

\begin{figure}
\label{sampleA}
\includegraphics[scale=0.30]{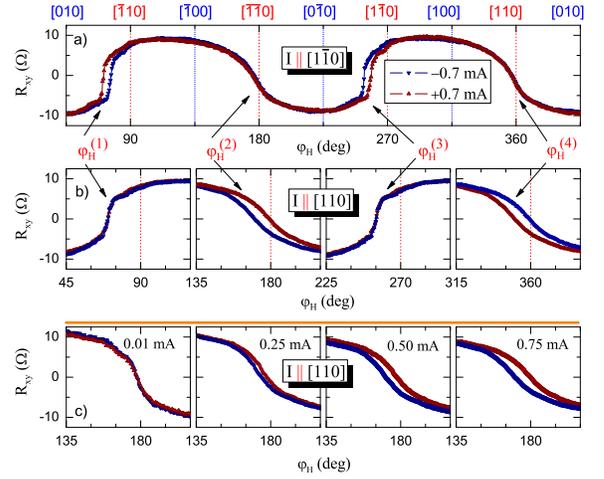}
\caption{{\bf Dependence of transverse anisotropic magnetoresistance on
current and field orientation} a,b) Transverse anisotropic magnetoresistance
$R_{xy}$ is plotted as a function of external field direction $\varphi_H$ for
$H=10$ mT and current $I=\pm 0.7$ mA in Sample A. The angles $\varphi_H^{(i)}$
mark magnetization switchings. In c) magnetization switching between
$[\bar{1}00]$ and $[0\bar{1}0]$ easy axes is plotted for several values of the
current.}
\end{figure}

In order to investigate interactions between the SO-generated magnetic field
and magnetic domains we have chosen (Ga,Mn)As, a p-type ferromagnetic
semiconductor\cite{ohno96,ohno98} with zinc blende crystalline structure
similar to GaAs. Ferromagnetic interactions in this material are
carrier-mediated\cite{dietl00,bercui01}. The total angular momentum of the
holes ${\bf J}$ couples to the magnetic moment ${\bf F}$ of Mn ions via
antiferromagnetic exchange ${\cal H}_{ex}=-A{\bf F}\cdot{\bf J}$. This
interaction leads to the ferromagnetic alignment of magnetic moments of Mn ions
and equilibrium polarization of hole spins. If additional, non-equilibrium spin
polarization of the holes $\langle\mathbf{J^E}\rangle$ is induced, the
interaction of the hole spins with magnetic moments of Mn ions allows one to
control ferromagnetism by manipulating $\mathbf{J}$. Magnetic properties of
(Ga,Mn)As are thus tightly related to the electronic properties of GaAs. For
example, strain-induced spin anisotropy of the hole energy dispersion is
largely responsible for the magnetic anisotropy in this material. (Ga,Mn)As,
epitaxially grown on (001) surface of GaAs, is compressively strained, which
results in magnetization $\mathbf M$ lying in the plane of the layer
perpendicular to the growth direction, with two easy axes along the [100] and
[010] crystallographic directions\cite{welp03,liu03}. Recently, control of
magnetization via strain modulation has been demonstrated\cite{overby08}. In
this paper we use SO-generated polarization $\langle\mathbf{J^E}\rangle$ to
manipulate ferromagnetism.

%\newpage

We report measurements on two samples fabricated from (Ga,Mn)As wafers with
different Mn concentrations. The devices were patterned into circular islands
with 8 non-magnetic Ohmic contacts, as shown in Fig.~\ref{sample}a and
discussed in Methods. In the presence of a strong external magnetic field
$\mathbf H$, the magnetization of the ferromagnetic island is aligned with the
field. For weak fields, however, the direction of magnetization is primarily
determined by magnetic anisotropy. As a small field ($5<H<20$ mT) is rotated in
the plane of the sample, the magnetization is re-aligned along the easy axis
closest to the field direction. Such rotation of magnetization by an external
field is demonstrated in Fig.~\ref{sampleA}. For the current $\mathbf
I||[1\bar{1}0]$, the measured $R_{xy}$ is positive for $\mathbf{M}||[100]$ and
negative for $\mathbf{M}||[010]$. Note that $R_{xy}$, and thus also the
magnetization,  switches direction when the direction of $\mathbf{H}$ is close
to the hard axes [110] and $[1\bar{1}0]$, confirming the cubic magnetic
anisotropy of our samples. The switching angles $\varphi_H=\angle\mathbf {H I}$
where $R_{xy}$ changes sign are denoted as $\varphi_H^{(i)}$ on the plot.

In the presence of both external and SO fields, we expect to see a combined
effect of  $\mathbf{H^{so}}+\mathbf{H}$ on the direction of magnetization. For
small currents (few $\mu$A) $H^{so}\approx0$, and  $R_{xy}$ does not depend on
the sign or the direction of the current. At large dc currents the value of
$\varphi_H^{(i)}$ becomes current-dependent and we define
$\Delta\varphi^{(i)}_H(I)=\varphi^{(i)}_H(I)-\varphi^{(i)}_H(-I)$.
Specifically, for $\mathbf I||[1\bar{1}0]$ the switching of magnetization
$[010]\rightarrow[\bar{1}00]$ occurs  for $I=+0.7$ mA at smaller
$\varphi^{(1)}_H$ than for $I=-0.7$ mA,  $\Delta\varphi^{(1)}_H<0$. For the
$[0\bar{1}0]\rightarrow[100]$ magnetization switching, the $I$-dependence of
switching angle is reversed, $\Delta\varphi^{(3)}_H>0$. There is no measurable
difference in switching angle for the $[\bar{1}00]\rightarrow[0\bar{1}0]$ and
$[100]\rightarrow[010]$ transitions ($\Delta\varphi^{(2,4)}_H\approx 0$). When
the current is rotated by 90$^\circ$ ($\mathbf{I}||[110]$), we observe
$\Delta\varphi_H^{(2)}>0$, $\Delta\varphi_H^{(4)}<0$, and
$\Delta\varphi_H^{(1,3)}\approx0$. In Fig.~\ref{sampleA}(c) we show that
$\Delta\varphi^{(2)}_H(I)$ decreases as current decreases and drops below
experimental resolution of $0.5^{\circ}$ at $I<50$ $\mu$A. Similar data is
obtained for Sample B, see Fig.~S4 in Supplementary Information.

The data can be qualitatively understood if we consider an additional
current-induced effective magnetic field $\mathbf{H^{eff}}$, as shown
schematically in Fig.~\ref{sample}b. When an external field $\mathbf H$ aligns
the magnetization along one of the hard axes, a small perpendicular field can
initiate magnetization switching. For $\mathbf I||[110]$, the effective field
$\mathbf{H^{eff}}||[\bar{1}10]$ aids the $[100]\rightarrow[010]$ magnetization
switching, while it hinders the $[\bar{1}00]\rightarrow[0\bar{1}0]$ switching.
For $\varphi_H^{(1)}\approx90^{\circ}$ and $\varphi_H^{(3)}\approx270^{\circ}$,
where $[010]\rightarrow[\bar{1}00]$ and $[0\bar{1}0]\rightarrow[100]$
magnetization transitions occur, $\mathbf{H^{eff}}||\mathbf{H}$ does not affect
the transition angle, $\Delta\varphi^{(2,4)}_H= 0$. For $\mathbf
I||[1\bar{1}0]$ the direction of the field $\mathbf{H^{eff}}||[110]$ is
reversed relative to the direction of the current, compared to the $\mathbf
I||[110]$ case. The symmetry of the measured $\mathbf{H^{eff}}$ with respect to
$\mathbf{I}$ coincides with the unique symmetry of the strain-related SO field,
Fig.~\ref{sample}(c).

The dependence of $\Delta\varphi_H^{(i)}$ on various magnetic fields and
current orientations is summarized in Fig.~\ref{current}(a,b). Assuming that
the angle of magnetization switching depends only on the total field
$\mathbf{H^{eff}+H}$, we can extract the magnitude $H^{\mathrm{eff}}$ and angle
$\theta=\angle\mathbf{IH^{eff}}$ from the measured $\Delta\varphi_H^{(i)}$,
thus reconstructing the whole vector $\mathbf{H^{eff}}$. Following a
geometrical construction depicted in Fig.~\ref{current}d and taking into
account that $\Delta\varphi_H^{(i)}$ is small, we find that
\[H^{\mathrm{eff}}\approx
H\sin(\Delta\varphi_H^{(i)}/2)/\sin(\theta-\varphi_H^{(i)}),\] and $\theta$ can
be found from the comparison of switching at two angles. We find that
$\theta\approx90^{\circ}$, or $\mathbf{H^{eff}}\bot\mathbf{I}$ for $I\|[110]$
and $I\|[1\bar{1}0]$. In order to further test our procedure we performed
similar experiments with small current $I=10$ $\mu$A but constant additional
magnetic field $\delta\mathbf H\bot\mathbf I$ playing the role of
$\mathbf{H^{eff}}$. The measured $\delta H(\Delta\varphi_H)$ coincides with the
applied $\delta H$ within the precision of our measurements. (see Fig.~S5 of
Supplementary Information).

\begin{figure}
\def\ffile{current}
\includegraphics[scale=0.30]{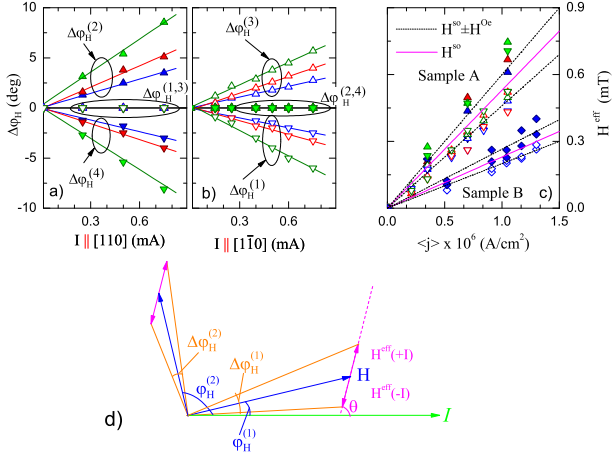}
\caption{{\bf Determination of current-induced effective SO magnetic field.}
a,b) Difference in switching angles for opposite current directions
$\Delta\varphi_H^{(i)}$ as a function of $I$ are plotted for Sample A for
different external fields $H$ for orthogonal current directions.  In c) the
measured effective field $H^{\mathrm{eff}}=H^{\mathrm{so}}\pm H^{\mathrm{Oe}}$
is plotted as a function of average current density $\langle j\rangle$ for
Sample A (triangles) and Sample B (diamonds). In d) we schematically show
different angles involved in determining $H^\mathrm{eff}$: $\varphi_H$ is the
angle between current $\mathbf{I}$ and external magnetic field $\mathbf{H}$;
$\Delta\varphi_H$ is the angle between total fields
$\mathbf{H}+\mathbf{H^{eff}}(+I)$ and $\mathbf{H}+\mathbf{H^{eff}}(-I)$, and
$\theta$ is the angle between $\mathbf{I}$ and $\mathbf{H^{eff}}(+I)$.}
\label{\ffile}
\end{figure}

\begin{figure}
\def\ffile{tamr-switch}
\includegraphics[scale=0.35]{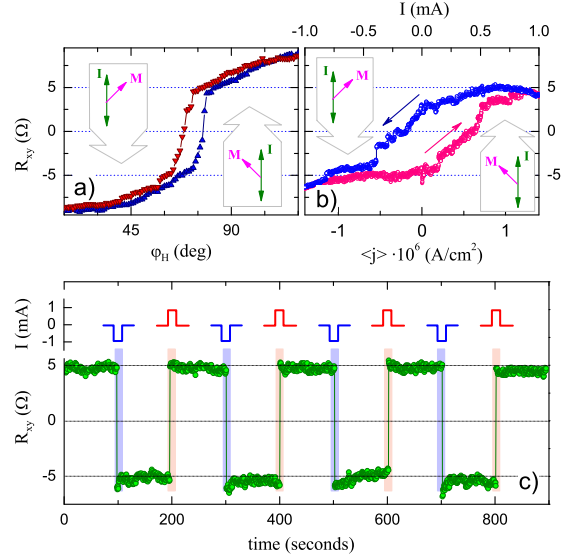}
\caption{{\bf Current-induced reversible magnetization switching} a)
$\varphi_H$-dependence of $R_{xy}$ near the $[010]\rightarrow[\bar{1}00]$
magnetization switching is plotted for $I=\pm0.7$ mA in Sample A for
$I\|[1\bar{1}0]$. b)$R_{xy}$ shows hysteresis as a function of current for a
fixed field $H=6$ mT applied at $\varphi_H=72^{\circ}$.  c) Magnetization
switches between $[010]$ and $[\bar{1}00]$ directions when alternating $\pm
1.0$ mA current pulses are applied. The pulses have 100ms duration and are
shown schematically above the data curve. $R_{xy}$ is measured with $I=10$
$\mu$A.}
\label{\ffile}
\end{figure}

In Fig.~\ref{current}(c), $H^{\mathrm{eff}}$ is plotted as a function of the
average current density $\langle j\rangle$ for both samples. There is a small
difference in the $H^{\mathrm{eff}}$ vs $\langle j\rangle$ dependence for
$\mathbf{I}\|[110]$ and $\mathbf{I}\|[1\bar{1}0]$. The difference can be
explained by considering the current-induced Oersted field
$H^{\mathrm{Oe}}\propto I$ in the metal contacts. The Oersted field is
localized under the pads, which constitutes only 7\% (2.5\%) of the total area
for samples A (B).  The Oersted field has the symmetry of the field depicted in
Fig.~\ref{sample}(d), and is added to or subtracted from the SO field,
depending on the current direction. Thus,
$H^{\mathrm{eff}}=H^{\mathrm{so}}+H^{\mathrm{Oe}}$ for $\mathbf{I}\|[110]$ and
$H^{\mathrm{eff}}=H^{\mathrm{so}}-H^{\mathrm{Oe}}$ for
$\mathbf{I}\|[1\bar{1}0]$. We estimate the fields to be as high as 0.6 mT under
the contacts at $I=1$ mA, which corresponds to $H^{Oe}\approx0.04$ mT (0.015
mT) averaged over the sample area for samples A (B). These estimates are
reasonably consistent with the measured values of 0.07 mT (0.03 mT). Finally,
we determine $H^{so}$ as an average of $H^{\mathrm{eff}}$ between the two
current directions. The SO field depends linearly on $j$, as expected for
strain-related SO interactions: $dH^{so}/dj=0.53\cdot10^{-9}$ and
$0.23\cdot10^{-9}$ T cm$^2$/A for samples A and B respectively.

We now compare the experimentally measured $H^{\mathrm{so}}$ with theoretically
calculated  effective SO field. In (Ga,Mn)As, the only term allowed by symmetry
that generates $H^{\mathrm{so}}$ linear in the electric current is the
$\Omega_\varepsilon$ term, which results in the directional dependence of
$\mathbf{H}^{{\mathrm{so}}}$ on $\mathbf{j}$ precisely as observed in
experiment. As for the magnitude of $H^{\mathrm{so}}$, for three-dimensional
$J=3/2$ holes we obtain
\[\mathbf{H^{so}(E})=\frac{eC\Delta\varepsilon}{g^*\mu_B}
\frac{(-38n_h\tau_h+18n_l\tau_l)}{217(n_h+n_l)}\cdot (E_x, -E_y,0),\] where
$\mathbf{E}$ is the electric field, $g^*$ is the Luttinger Land\'e factor for
holes, $\mu_B$ is the Bohr magneton, and $n_{h,l}$ and $\tau_{h,l}$ are
densities and lifetimes for the heavy (h) and light (l) holes. Detailed
derivation of $H^{\mathrm{so}}$ is given in Supplementary Information. Using
this result, we estimate $dH^{so}/dj=0.6\cdot10^{-9}$ T cm$^2$/A assuming
$n_h=n\gg n_l$ and $\tau_h=m_h/(e^2\rho n)$, where $\rho$ is the resistivity
measured experimentally, and using $\Delta\varepsilon=10^{-3}$, $n=2\cdot
10^{20}$ cm$^{-3}$. The agreement between theory and experiment is excellent.
It is important to note, though,  that we used GaAs band
parameters\cite{chantis08} $m_h=0.4 m_0$, where $m_0$ is the free electron
mass, $g^*=1.2$ and $C=2.1$ eV$\cdot$\AA. While the corresponding parameters
for (Ga,Mn)As are not known, the use of GaAs parameters appears reasonable. We
note, for example, that GaAs parameters adequately described tunnelling
anisotropic magnetoresistance in recent experiments\cite{elsen07}.

Finally, we demonstrate that the current-induced effective SO field $H^{so}$ is
sufficient to reversibly manipulate the direction of magnetization. In
Fig.~\ref{tamr-switch}a we plot the $\varphi_H$-dependence of $R_{xy}$ for
Sample A, showing the $[010]\rightarrow[\bar{1}00]$ magnetization switching. If
we fix $H=6$ mT at $\varphi_H=72^{\circ}$, $R_{xy}$ forms a hysteresis loop as
current is swept between $\pm1$ mA. $R_{xy}$ is changing between $\pm5$
$\Omega$, indicating that $\mathbf M$ is switching between $[010]$ and
$[\bar{1}00]$ directions. Short (100 msec) 1 mA current pulses of alternating
polarity are sufficient to permanently rotate the  direction of magnetization.
The device thus performs as a non-volatile memory cell, with two states encoded
in the magnetization direction, the direction being controlled by the
unpolarized current passing through the device. The device can be potentially
operated as a 4-state memory cell if both $[110]$ and $[\bar{1}10]$ directions
can be used to inject current. We find that we can reversibly switch the
magnetization with currents as low as 0.5 mA (current densities $7\cdot10^5$
A/cm$^2$), an order of magnitude smaller than by polarized current injection in
ferromagnetic metals\cite{slonczewski96,berger96,myers99}, and just a few times
larger than by externally polarized current injection in ferromagnetic
semiconductors\cite{chiba04}.

{\bf Methods}

The (Ga,Mn)As wafers were grown by molecular beam epitaxy at 265 $^\circ$C and
subsequently annealed at 280 $^\circ$C for 1 hour in nitrogen atmosphere.
Sample A was fabricated from 15-nm thick epilayer with 6\%Mn, and Sample B from
10-nm epilayer with 7\% Mn. Both wafers have Curie temperature $T_c\approx 80$
K. The devices were patterned into 6 and 10 $\mu$m-diameter circular islands in
order to decrease domain pinning. Cr/Zn/Au (5nm/10nm/300nm) Ohmic contacts were
thermally evaporated. All measurements were performed in a variable temperature
cryostat at $T=40$ K for Sample A and at 25 K for Sample B, well below the
temperature of (Ga,Mn)As-specific cubic-to-uniaxial magnetic anisotropy
transitions\cite{chiba08}, which has been measured to be at 60 K and 50 K for
the two wafers. Temperature rise for the largest currents used in the reported
experiments was measured to be $<3$ K.

Transverse anisotropic magnetoresistance $R_{xy}=V_y/I_x$ is measured using the
four-probe technique, which insures that possible interfacial resistances,
e.g., those related to the antiferromagnetic ordering in the Cr wetting
layer\cite{smit88}, do not contribute to the measured $R_{xy}$. The DC current
$I_x$ was applied either along [110] (contacts 4-8 in Fig.~\ref{sample}a) or
along $[1\bar 10]$ (contacts 2-6) direction. Transverse voltage was measured in
the Hall configuration, e.g., between contacts 2-6 for $I_x \| [110]$.  To
ensure uniform magnetization of the island, magnetic field was ramped to 0.5 T
after adjusting of the current at the beginning of each field rotation scan. We
monitor $V_x$ between different contact sets (e.g. 1-7, 4-6 and 3-5) to confirm
the uniformity of magnetization within the island.

In order to determine the direction of magnetization $\mathbf{M}$, we use the
dependence of $R_{xy}$ on magnetization\cite{tang03}:
\[R_{xy}=\Delta\rho\sin\varphi_M\cos\varphi_M,\] where
$\Delta\rho=\rho_\|-\rho_\bot$, $\rho_\|<\rho_\bot$ are the resistivities for
magnetization oriented parallel and perpendicular to the current, and
$\varphi_M=\angle\mathbf {M I}$ is an angle between magnetization and current.
In a circular sample the current distribution is non-uniform and the angle
between the magnetization and the local current density varies throughout the
sample. However, the resulting transverse AMR depends only on $\varphi_M$. For
the current-to-current-density conversion, we model our sample as a perfect
disc with two point contacts across the diameter. The average current density
in the direction of current injection is $\langle j\rangle=2I/(\pi ad)$, where
$a$ is the disk radius and $d$ is the (Ga,Mn)As layer thickness. In a real
sample the length of contact overlap with (Ga,Mn)As insures that $j$ changes by
less than factor of 3 throughout the sample. A detailed discussion of the
current distribution and of measurements of Joule heating can be found in
Supplementary Information.

\bibliography{rohi}

\onecolumngrid
\newpage

\begin{center}
\textbf{\Large Supplementary Information} \\
\vspace{0.2in} \textsc{Evidence for the reversible control of magnetization in a ferromagnetic material via spin-orbit magnetic field}\\
{\it A. Chernyshov, M. Overby, X. Liu, J. K. Furdyna, Y. Lyanda-Geller, and L.
P. Rokhinson}
\end{center}

\section{Joule heating}

(Ga,Mn)As is a magnetic semiconductor with strong temperature dependence of
resistivity, see Fig.~\ref{t-dep}(a). The enhancement of resistivity at 80 K is
due to the enhancement of spin scattering in the vicinity of the Curie
temperature $T_C$. Inelastic scattering length in these materials is just a few
tenths of nm, and we expect holes to be in thermal equilibrium with the
lattice\cite{rokhinson07}. Thus resistivity can be used to measure the
temperature of the sample.

\begin{figure}[h]
\def\ffile{t-dep}
\center
\includegraphics[scale=0.4]{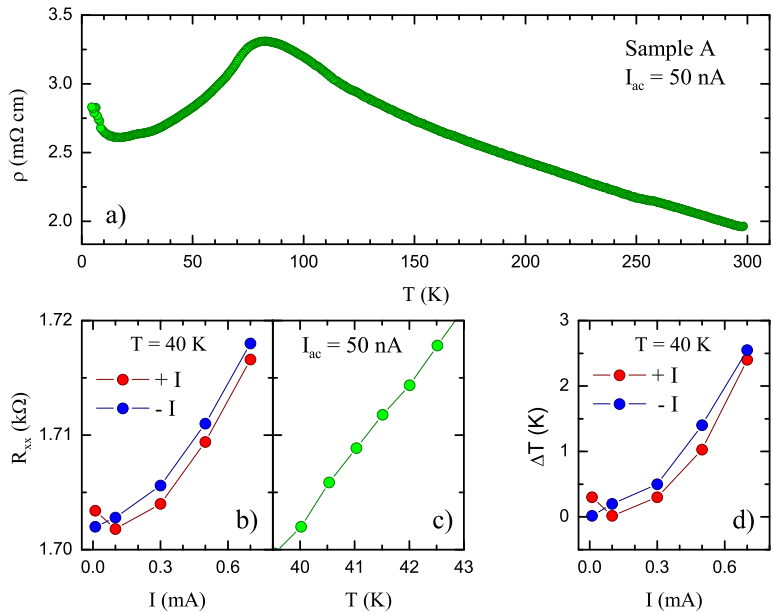}
\caption{
\textbf{Current-induced heating} a) Temperature dependence of resistivity for
sample A; b) current and c) temperature dependence of sample resistance in the
vicinity of 40 K; d) sample heating as a function of dc current.}\label{\ffile}
\end{figure}

In Fig.~\ref{t-dep}(b,c) we plot temperature and current dependences of the
sample resistance in the vicinity of 40 K. This data is combined in (d), where
the sample temperature change $\Delta T$ due to Joule heating is plotted as a
function of dc current. The maximum temperature rise does not exceed 3 K at $I
= 0.7$ mA in our experiments. This small heating ensures that the sample
temperature stays well below the Curie temperature ($\approx 80$ K) and the
(Ga,Mn)-specific cubic-to-uniaxial magnetic anisotropy transition ($\approx 60$
K for sample A and $\approx 50$ K for sample B) when experiments are performed
at 40 K and 25 K for samples A and B, respectively. Observation of different
angles for magnetization switching for $+I$ and $-I$ (Fig.~2) further confirms
that heating is not responsible for the reported effects (Joule heating is
$\propto J^2$ and does not depend on the current direction).

\section{Current distribution in circular samples}

Magnetization-dependent scattering in (Ga,Mn)As results in an anisotropic
correction to the resistivity tensor $\hat{\rho}$ which depends on the angle
$\varphi_m$ between magnetization $\mathbf{M}$ and local current density
$\mathbf{j}$ \cite{tang03}:
\begin{eqnarray}
\label{AMR}
\rho_{xx}=\rho_{\bot}+(\rho_{\|}-\rho_{\bot})\cos^2(\varphi_{m})\text{,}\nonumber\\
\rho_{xy}=(\rho_{\|}-\rho_{\bot})\sin(\varphi_{m})\cos(\varphi_{m}),
\end{eqnarray}
where $\rho_{\|}$ ($\rho_{\bot}$) are the resistivities for $\mathbf{j||M}$
($\mathbf{j\bot M}$), and we assumed that both $\mathbf{j}$ and $\mathbf{M}$
lie within the plane of the sample. The off-diagonal resistivity (transverse
anisotropic magnetoresistance) $\rho_{xy}$ can be non-zero even in the absence
of the external magnetic field. The difference
$(\rho_{||}-\rho_{\bot})/\rho_{\bot}\approx 0.01$ and we first calculate the
local potential $\phi_0(x,y)$ inside the sample by approximating it as a disk
of radius $a$ and thickness $d$ with isotropic resistivity
$\rho_{0}=(\rho_{\|}+\rho_{\bot})/2$:
\begin{equation}\label{Pot}
\phi_0=\frac{\rho_{0}I}{\pi d}\ln\Big[\frac{(a-x)^2+y^2}{(a+x)^2+y^2}\Big],
\end{equation}
where current $I$ is injected along the $\hat{x}$-axis. Current density
$\mathbf{j}=\nabla\phi_0/\rho_0$ is plotted in Fig.~\ref{current1}(a). Metal
contacts have a radius of $\approx0.5$ $\mu$m in our samples, which limits the
current density near the current injection regions. Integrating $j$ over the
sample area we find average current density
\begin{equation}
\langle j_x\rangle=\frac{2I}{\pi a d},\ \langle j_y\rangle=0.
\end{equation}

\begin{figure}[h]
\def\ffile{current1}
\vspace{.5in}
\includegraphics[scale=0.5]{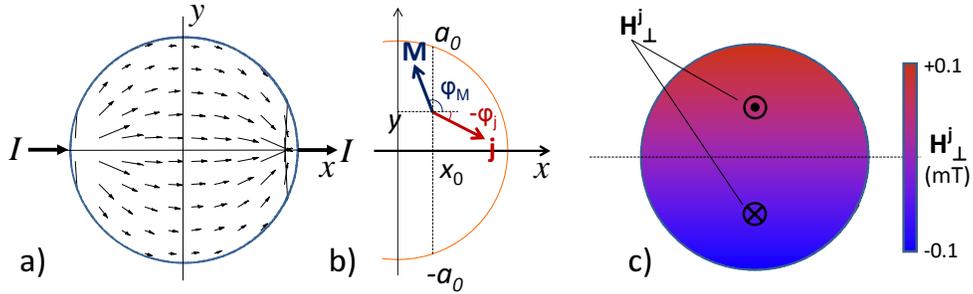}
\caption{\textbf{Current distribution} a) Vector plot of local current density $\mathbf{j}(x,y)$ distribution
in the sample; b) angles between $\mathbf{j}(x,y)$, magnetization $\mathbf{M}$
and current $\mathbf{I}\|\hat{x}$ are defined; c) Color map plot of Oersted
field ($H^j_{\bot}$) distribution in a disk-shaped sample. }
\label{\ffile}
\end{figure}

We find the transverse voltage $V_y$ as a correction to the $\phi_0$ potential
due to the anisotropic resistivity $\rho_{||}-\rho_{\bot}\neq0$:
\begin{equation}\label{Vy}
V_{y}(x_0)=\int_{-a_0}^{a_0}\big[-\rho_{xy}\cdot j_x(y)+\rho_{xx}\cdot
j_y(y)\big] dy.
\end{equation}
The current distribution is non-uniform, and the local electric field depends
on the total angle $\varphi_m=\varphi_M-\varphi_j$, where
$\varphi_M=\mathbf{\widehat{MI}}$ and $\varphi_j=\mathbf{\widehat{jI}}$, see
Fig.~\ref{current1}(b). This integral can be evaluated analytically, and the
transverse anisotropic magnetoresistance (AMR) $R_{xy}$ is found to be the same
as for an isotropic current flow, independent of the distance $x_0$ of the
voltage contacts from the center of the disk:
\begin{equation}\label{rxy}
R_{xy}=V_y/I = (\rho_{\|}-\rho_{\bot})\cos(\varphi_M)\sin(\varphi_M).
\end{equation}
The magnetization angle $\varphi_M$ can therefore be directly calculated from
the measured transverse resistance $R_{xy}$.

\section{Current-generated Oersted magnetic fields}

In this section we estimate conventional current-generate magnetic fields in
our device that are not related to spin-orbit interactions. There are two
contributions to the Oersted magnetic fields: a magnetic field due to
non-uniform current distribution within the sample, and a field generated by
high currents in the vicinity of the metal contacts.

We can calculate the Oersted field inside (Ga,Mn)As by using the Biot-Savart
formula:
\begin{equation}\label{Hz}
\mathbf{H}=\frac{\mu_0}{4\pi}\int\frac{\mathbf{j\times\hat{r}}}{r^2}dV,
\end{equation}
where $\mu_0$ is the permeability of free space, and the integral is taken over
the volume of the disk. The most significant $H_{\bot}^{j}$ normal component of
the field is shown in Fig.~\ref{current}(c). The largest $H_{\bot}^{j}\leq1$
Oe, which is negligible compared to the 2000 Oe anisotropy field that keeps the
magnetization in-plane.

\begin{figure}[h]
\def\ffile{contacts}
\includegraphics[scale=0.5]{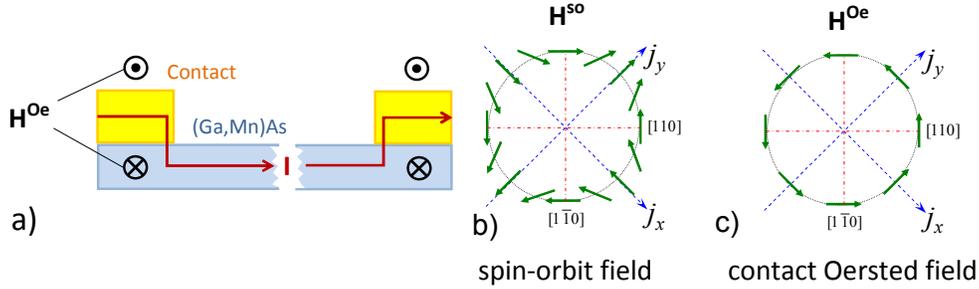}
%\vspace{-2in}
\caption{
\textbf{Oersted field} a) schematic illustration of the origin of the in-plane
Oersted field $H^{Oe}$ under gold contact pads; b,c) symmetry of $H^{so}$ and
$H^{Oe}$ fields.}\label{\ffile}
\end{figure}

The second contribution to the Oersted field originates from contact pads, see
Fig.~\ref{contacts}. The conductivity of gold contacts is much higher that of
(Ga,Mn)As, and the current flows predominantly through the metal within contact
regions, thus generating both in-plane ($H_{\|}^{Oe}$) and out-of plane
($H_{\bot}^{Oe}$) magnetic fields in (Ga,Mn)As underneath and at the edges of
the contact pads. The maximum value of the field can be estimated as
$H_{\bot}^{Oe}\approx H_{||}^{Oe}=\mu_0 I/2w$, where $I$ is the total current
and $w=1\ \mu$m is the width of the contact pad. This field can be as high as 6
Oe for $I = 1$ mA. The field is localized under the pads, which constitute only
1/12th of the sample area.

The $H_{\bot}^{Oe}$ field does not induce in-plane magnetization rotation. The
$H_{\|}^{Oe}$ field and the effective spin-orbit field have different
symmetries with respect to the current rotation, see Fig.~\ref{current}(b,c),
and thus can be experimentally distinguished. The two fields point in the same
direction for $\mathbf{I}||[110]$, but in the opposite direction for the
current rotated by $90^{\circ}$, $\mathbf{I}||[1\bar{1}0]$. Experimentally, we
observe an effective field which corresponds to the symmetry of the SO
effective field. However, there is a small difference in the slopes of
$\Delta\phi_H$ vs $I$ curves for the two orthogonal current directions,
Fig.~3(a,b), because the contact field is added to the SO field for
$\mathbf{I}||[110]$ and subtracted from SO field for $\mathbf{I}||[1\bar{1}0]$.
Both fields $\propto I$. From the ratio of the slopes ($\approx1.2$) we can
calculate the strength of the contact field, $H_{\bot}^{Oe}\approx0.1 H^{so}$.
This experimentally found ratio is consistent with the above estimate if we
average the contact Oersted field over the sample area.

%\newpage

\section{Dependence of transverse anisotropic magnetoresistance on
current and field orientation}

\begin{figure}[h]
\def\ffile{SampleB}
\includegraphics[scale=0.4]{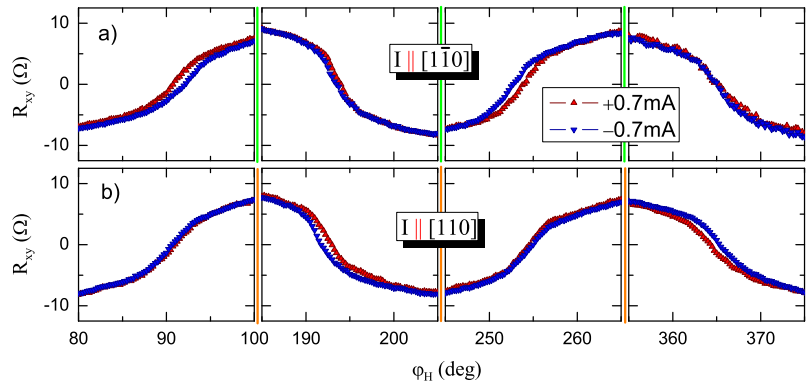}
\caption{
\textbf{Dependence of transverse anisotropic magnetoresistance on current and
field orientation for Sample B.} Transverse anisotropic magnetoresistance
$R_{xy}$ is plotted for current $\mathbf I||[1\bar{1}0]$ (a) and $\mathbf
I||[110]$ (b) for $I=\pm 0.75$ mA with constant magnetic field $H=20$ mT as a
function of field angle $\varphi_H$. }
\label{\ffile}
\end{figure}

In order to test the procedure of the effective field mapping, we performed
control experiments where a small constant external magnetic field
$\delta\mathbf H\bot\mathbf I$ was playing the role of spin-orbit field. In
these experiments the current was reduced to $I=10$ $\mu$A. The results, shown
in Fig.~\ref{Beff}, are quantitatively similar to the effect of the spin-orbit
field.

\begin{figure}[h]
\def\ffile{Beff}
\includegraphics[scale=0.4]{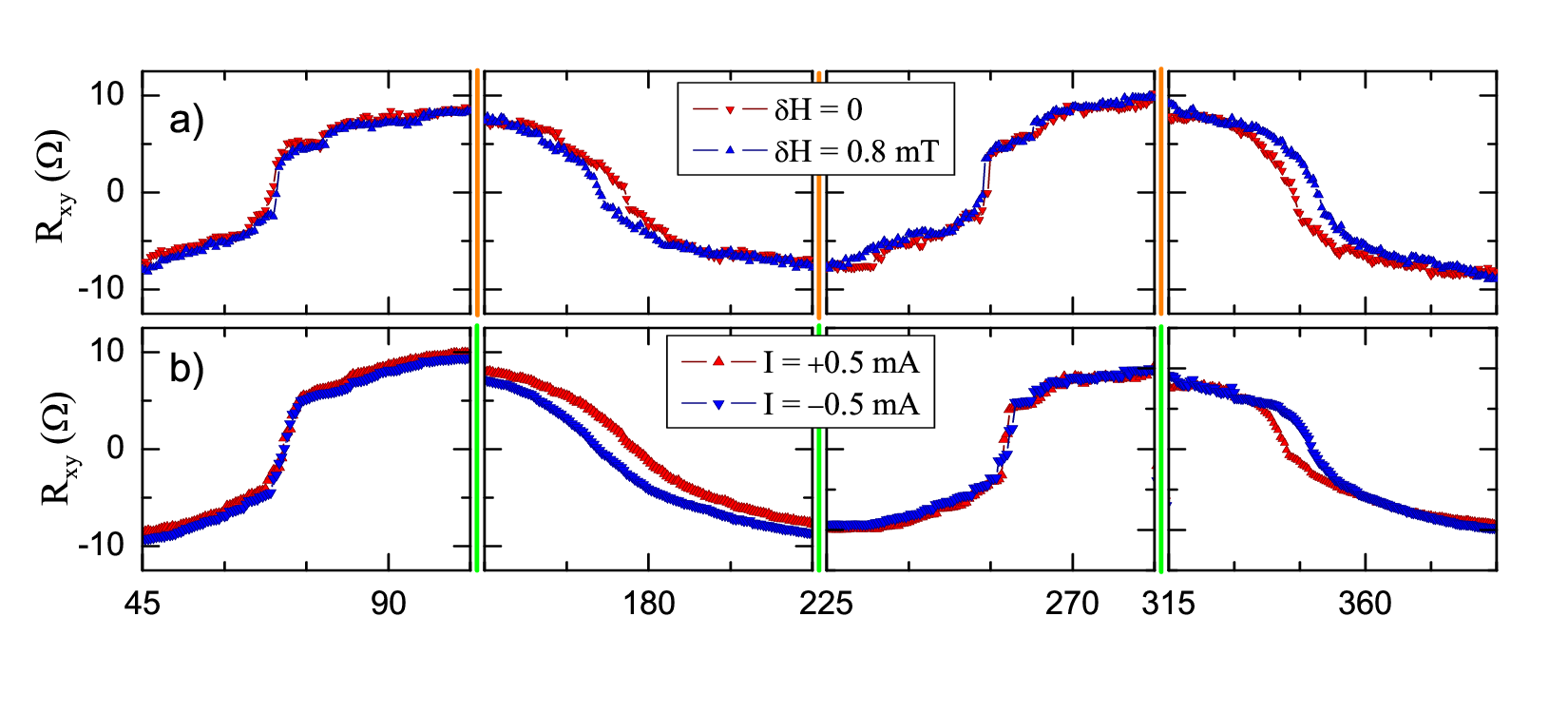}
\caption{
\textbf{Control experiment with additional external field} a) $R_{xy}$ is
plotted for Sample A with $\mathbf{H_{total}}=\mathbf{\delta H} + \mathbf{H}$,
where $\mathbf\delta H\bot\mathbf I$ and $I=10$ $\mu$A, $\delta H=0, 0.8$ mT.
For comparison, in b) similar data are plotted for $\delta H=0$ but $I=\pm0.5$
mA. }
\label{\ffile}
\end{figure}

\section{Calculation of the effective spin-orbit field induced by the electric
current}

Manipulation of localized spins, electronic, nuclear or ionic, can be achieved
via manipulation of free carrier spins. The free carrier spins can be
manipulated by the external magnetic field, by the Oersted magnetic field of
the current, and by the electric current via intrinsic spin-orbit interactions.
The intrinsic spin-orbit interactions arise in crystalline systems, in which
axial vectors, such as spin polarization, and polar vectors, such as the
electric current, behave equivalently with respect to the symmetry
trasformations of a crystal. The crystal symmetry then allows the
transformation of the electric current into a spin polarization of charge
carriers. In this work, Mn ions of the ferromagnetic semiconductor (Ga,Mn)As,
and thus its ferromagnetic properties, are affected by the electric current via
the intrinsic spin-orbit interactions.

In (Ga,Mn)As, charge carriers are holes with an angular momentum ${\mathbf
J}=3/2$. In contrast to electron systems, the hole system is defined by a very
strong coupling of the total angular momentum $\mathbf{J}$ to the hole momentum
$\mathbf{p}$, which includes both terms quadratic in $\mathbf{p}$ and
independent of $\mathbf{p}$. These terms are quadratic in $\mathbf{J}$, and
they are not present for electrons with spin 1/2. The Luttinger-Pikus
Hamiltonian quadratic in $\mathbf{J}$ is\cite{birpikus74}
\begin{equation}
{\cal H}_h= A_0 p^2+A_1\sum_{i}J_i^2p_i^2 +A_2\sum_{i,j\ne i}J_iJ_jp_ip_j+
B_1\sum_{i}\varepsilon_{ii}J_i^2 +B_2\sum_{i,j\ne i}J_iJ_j\varepsilon_{ij},
\end{equation}
where $i,j=x,y,z$. Despite the presence of a very strong spin-orbit coupling,
which leads to the spectral splitting of holes into two pairs of states, this
Hamiltonian on its own cannot result in a spin polarization of holes induced by
the electric current. Terms capable of generating spin polarization in systems
characterized by the absence of center of symmetry in the crystal and by a
corresponding additional lowering of the crystalline symmetry in the presence
of strain, read
\begin{equation}
{\cal H}^{\prime}= \gamma_v \sum_{i}J_i p_i(p_{i+1}^2-p_{i+2}^2)+ C\sum_{i}[
J_ip_i(\varepsilon_{i+1,i+1}-\varepsilon_{i+2,i+2}) +
(J_ip_{i+1}-J_{i+1}p_i)\varepsilon_{i,i+1}],
\end{equation}
where cyclic permutation of indices is implied. The first term is cubic in the
hole momentum, and it can lead only to the polarization of hole spins cubic in
the electric current (and only when the current direction is away from the high
symmetry axes). For effects linear in electric current this term is only
relevant insofar as it contributes to the spin relaxation of the holes. The
third term contains off-diagonal components of the strain tensor, and is
negligible in (Ga,Mn)As crystals under study. In this system, strain originates
from doping by Mn ions, and constitutes tension along the growth axis
$z||[001]$ defined by the component $\varepsilon_{zz}$ and
$\Delta\varepsilon=\varepsilon_{zz}-\varepsilon_{xx}=\varepsilon_{zz}-\varepsilon_{yy}$.
Thus only the second term results in a current-induced spin polarization. The
symmetry of the corresponding effective field,
$\mathbf{\Omega}(\mathbf{p})=C\Delta\varepsilon(p_x,-p_y,0)$, depends markedly
on the crystallographic orientation. When an electric field is applied, the
direction of the generated hole spin polarization with respect to the
orientation of the electric current is the same as the direction of the SO
effective field with respect to the hole momentum. Such peculiar symmetry
differs from the symmetry of the Oersted magnetic field, and thus allows one to
distinguish between these effects.

We consider now the approximation linear in strain, when only the
strain-dependent term proportional to $C$ is taken into account, and
strain-dependent terms in ${\cal H}_h$ are omitted. In this case the hole
spectrum given by ${\cal H}_h$ splits into heavy $(h)$ and light $(l)$ hole
branches. The mechanism of generation a spin polarization by the effective SO
field in the presence of an electric current is simply a shift in the
distribution functions for heavy and light holes in momentum space. In contrast
low symmetry electron systems\cite{aronov91}, where spin polarization is
associated entirely with the relaxation of spins, in case of holes the spin
relaxation occurs on the time scale of momentum relaxation and plays no role in
the current-induced spin polarization. At low temperatures the hole angular
momentum density is given by
\begin{equation}
\label{eq2}
\langle J_{i}^{(\mathbf{E})}\rangle= (-1)^i\frac{eE_{i}C
\Delta\varepsilon}{E_F}\Big(\frac{-38}{35}n_h\tau_h +
\frac{18}{35}n_l\tau_l\Big),
\end{equation}
where $i=1,2$ correspond to principal axes $x$ and $y$, characteristic times
$\tau_{h,l}$ are defined by mobilities of holes in the corresponding bands, and
$n_{h(l)}$ are densities of holes in these bands. At room temperatures $E_F$ in
the denominator is to be replaced by $3/2k_BT$, $T$ being the lattice
temperature and $k_B$ the Boltzman constant. Estimates show that the negative
term in brackets of Eq.~\ref{eq2} is dominant.

We note that in the case of very strong deformations the spin relaxation of
holes occurs on the times scale longer than that of momentum relaxation. Then
simple shift of hole distribution functions in momentum space is no longer
sufficient for generating spin polarization by current, and the mechanism of
the effect becomes analogous to that for electrons\cite{aronov91}. We will
present the results for hole spin polarization generated by electric current at
arbitrary value of strain elsewhere.

The spin polarization given by Eq.~\ref{eq2} leads to an effective magnetic
field acting on the Mn ions. In order to calculate what external magnetic field
would result in the same polarization as that generated by the current, we
calculate the average spin density induced by an external magnetic field:
\begin{equation}
\langle J_i^{(\mathbf{H)}}\rangle=\frac{31g^*\mu_B H(n_h+n_l)}{5E_F}
\end{equation}
The ratio of polarizations $\langle J_i^{(\mathbf{E})}\rangle$ and $\langle
J_i^{(\mathbf{H)}}\rangle$ gives the electric field polarization measured in
units of magnetic field. We note that while the SO field affects Mn ions only
via the exchange interaction, the Oersted or the external magnetic field also
acts on the ions directly. However, the magnitude of the exchange interaction,
$A=-5$ meV is quite large, making the exchange interaction dominant. We will
therefore omit the discussion of direct polarization of Mn by external fields.

\end{document}